\documentclass[onecolumn,showpacs,preprintnumbers,amsmath,amssymb,APSl,nofootinbib,prd]{revtex4-1}
\usepackage{bm}
\usepackage{color,graphicx,graphics}
\usepackage[all]{xy}
\usepackage{epsfig,subfigure}
\usepackage{latexsym,amssymb,amsmath,amsfonts} 
\usepackage[english]{babel} 
\usepackage[OT1]{fontenc}
\usepackage[latin1]{inputenc}
\usepackage{makeidx}
\usepackage[normalem]{ulem}

\definecolor{red}{rgb}{1,0,0}

\def\<{\leftarrow}
\def\>{\rightarrow}

\def\({\left(}  \def\){\right)}

\newcommand{\bi}{\begin{itemize}} 		\newcommand{\ei}{\end{itemize}}
\newcommand{\benu}{\begin{enumerate}} \newcommand{\enu}{\end{enumerate}}
\newcommand{\bd}{\begin{dinglist}{0}}     \newcommand{\ed}{\end{dinglist}}
\newcommand{\bfig}{\begin{figure}[htbp]}  \newcommand{\efig}{\end{figure}}
       			
\newcommand{\bc}{\begin{center}} 	\newcommand{\ec}{\end{center}}
\newcommand{\be}{\begin{equation}} 	\newcommand{\ee}{\end{equation}}
\newcommand{\bsub}{\begin{subequations}}  \newcommand{\esub}{\end{subequations}}
\newcommand{\ben}{\begin{eqnarray}} 	\newcommand{\een}{\end{eqnarray}}
\newcommand{\ba}[1]{\begin{array}{#1}} 	\newcommand{\ea}{\end{array}}
\newcommand{\bea}{\begin{equation}\begin{array}{rcl}} \newcommand{\eea}{\end{array}\end{equation}}

\begin{document}
\title{Conformal Metric-Affine Gravities}
 
\author{Gonzalo J. Olmo$^{a,b}$}\email{gonzalo.olmo@uv.es}
\author{Emanuele Orazi$^{c}$}\email{orazi.emanuele@gmail.com}
\author{Gianfranco Pradisi$^{d}$}\email{gianfranco.pradisi@roma2.infn.it}
\affiliation{$^{a}$Departamento de F\'{i}sica Te\'{o}rica and IFIC, Centro Mixto Universidad de Valencia - CSIC.
	Universidad de Valencia, Burjassot-46100, Valencia, Spain}
\affiliation{$^{b}$Departamento de F\'isica, Universidade Federal do Cear\'{a} (UFC), Campus do Pici, Fortaleza - CE, C.P. 6030, 60455-760, Brazil}	
	
\affiliation{$^{c}$Escola de Ci\^encia e Tecnologia and International Institute of Physics,
Federal University of Rio Grande do Norte,
Campus Universit\'ario-Lagoa Nova, Natal-RN 59078-970, Brazil}
\affiliation{$^{d}$University of Rome ``Tor Vergata'' and INFN, Sezione di Roma ``Tor Vergata", via della Ricerca Scientifica 1, 00133 Roma, Italy}
\vskip 2cm

\date{\today}

\begin{abstract}
We revisit the gauge symmetry related to integrable projective transformations in metric-affine formalism, identifying the gauge field of the Weyl (conformal) symmetry as a dynamical component of the affine connection.  In particular, we show how to include the local scaling symmetry as a gauge symmetry of a large class of geometric gravity theories, introducing a compensator dilaton field that naturally gives rise to a St\"uckelberg sector where a spontaneous breaking mechanism of the conformal symmetry is at work to generate a mass scale for the gauge field.  For Ricci-based gravities that include, among others,  General Relativity, $f(R)$ and $f(R,R_{\mu\nu}R^{\mu\nu})$ theories and  the EiBI model, we prove that the on-shell gauge vector associated to the scaling symmetry can be identified with the torsion vector, thus recovering and generalizing conformal invariant theories in the Riemann-Cartan formalism, already present in the literature.
\end{abstract}

\maketitle


\section*{INTRODUCTION}


The great deal of interest in possible gravity theories sharing conformal symmetry is not recent. The first pioneering attempt to include conformal invariance in a geometrical formulation of gravitation can be traced back to the original Weyl proposal for a unified theory of gravitation and electromagnetism \cite{Weyl:1918ib}. The idea was to promote the local scaling\footnote{In literature, this also indicated as conformal scaling of the metric or Weyl transformation of the metric.} of the metric
\be
g^\prime_{\mu\nu} = e^{2\alpha(x)}g_{\mu\nu}\label{eq:WeylTransf}
\ee
to a gauge symmetry by introducing an associated gauge vector field represented by the so-called Weyl vector ${\cal W}_\mu$. It gauge transforms according to ${\cal W}'_\mu = {\cal W}_\mu + \partial_\mu\alpha(x)$,  in such a way that the metricity condition $\nabla_\mu g_{\nu\rho} = 0$ can be replaced by its gauge covariant version
\be
\nabla_\mu g_{\nu\rho} = 2 {\cal W}_\mu g_{\nu\rho}\,.\label{eq:NonMetrWeyl}
\ee
As a consequence, the underlying geometry is described by a non-Riemannian affine connection
\be
\Gamma_{(w)}^\mu{}_{\nu\rho} = \left\{
					\begin{array}{c}
					\mu\\
					\nu\rho
					\end{array}
				   \right\}
- \delta^\mu_\nu{\cal W}_\rho - \delta^\mu_\rho{\cal W}_\nu + g_{\nu\rho}{\cal W}^\mu\,,\label{eq:WeylAffineConn}
\ee
that allows to define a conformal invariant curvature $R_{(w)}^\alpha{}_{\beta\mu\nu} = 2\left(\partial_{[\mu} \Gamma_{(w)}^{\alpha}{}_{\nu]\alpha\beta} + \Gamma_{(w)}^\alpha{}_{[\mu\lambda}\Gamma_{(w)}^\lambda{}_{\nu]\beta} \right)$\,.

A distinctive feature of the Weyl geometry is that the affine connection in eq. \eqref{eq:WeylAffineConn} remains invariant under a Weyl transformation since the Weyl vector transformation compensates the one of the Christoffel symbol. Moreover, the Weyl geometry avoids the introduction of any torsional degrees of freedom since $\Gamma_{w}{}^\mu{}_{[\nu\rho]}=0$. Despite its success in introducing the world to the invention of gauge theories \cite{Weyl:1919fi, book:Oraif}, Weyl proposal was soon rejected in its original form. The main reason can be ascribed to the dependence of the measure of length on spacetime paths, that forbids the identification of a reference length scale.  Also, the gauge group can be identified with $\mathbb{R}^+$ that, being non-compact, is not suited to a description of electromagnetism.  A modern description of the Weyl space structure in terms of the nonmetricity tensor $Q_{\mu\nu\rho}$, based on the identification
\be
Q_{\mu\nu\rho} =  2 {\cal W}_\mu g_{\nu\rho}\,,
\ee
where the Weyl vector  ${\cal W}_\mu = \frac18 g^{\nu\rho} Q_{\mu\nu\rho} =  \frac18  Q_\mu$ is traded by the trace of nonmetricity, leads to gauge transformations of the nonmetricity of the form
\be
Q'_{\mu\nu\rho} =  e^{2\alpha}\left(Q_{\mu\nu\rho} + 2\partial_\mu\alpha\,g_{\nu\rho}\right)\,
\ee
and it is still plagued by the same issues as the Weyl proposal.

Following the original idea of \cite{Obukhov:1982zn}, it is possible to recover metric compatibility at the price of introducing torsion, inevitably moving to the more familiar realm of Einstein-Cartan geometry \cite{sciama:1962, Kibble:1961ba, Hehl:1974cn} (for a review, see e.g. \cite{Hehl:1976kj}). In practice, the metricity condition $\hat\nabla_\mu g_{\nu\rho} = 0$ can be restored from \eqref{eq:NonMetrWeyl} if a new  covariant derivative $\hat\nabla$ is introduced in terms of a connection that comprises the Weyl vector, $\hat\Gamma^\mu{}_{\nu\rho} =  \Gamma^\mu{}_{\nu\rho} + \delta^\mu_\nu{\cal W}_\rho$\footnote{
It is important not to confuse the introduction of torsion given here with the more general solution  $\hat\Gamma^\mu{}_{\nu\rho} =\left\{
					\begin{array}{c}
					\mu\\
					\nu\rho
					\end{array}
				   \right\}
+  K^\mu{}_{\nu\rho}$ in absence of nonmetricity,  where $K^\mu{}_{\nu\rho}$ is the contortion tensor. Having in mind a metric-affine approach, the metricity condition does not hold apriori off-shell, so that all we can say is that the connection can be split into its symmetric and antisymmetric part with respect to the lower indices.  Of course, once the connection satisfies the metricity condition, the tensorial part of the connection necessarily coincide with some contortion tensor.
}
. It follows that the additional term in the connection gives rise to torsional degrees of freedom $S^\mu{}_{\nu\rho} =  \delta^\mu_{[\nu}{\cal W}_{\rho]}$ and leads to an induced conformal transformation that acts on the torsion vector $S_\rho\equiv \delta^\nu_\mu S^\mu{}_{\nu\rho}=\frac32{\cal W}_\rho$  as $S^\prime_\mu = S_\mu + \frac32 \partial_\mu \alpha$. In other words, the gauge field associated to the conformal invariance is now understood as the torsion vector. As a consequence, contrary to the Weyl proposal,  the connection is no more invariant under a conformal transformation but gauge transforms according to the rule
\be
\hat\Gamma'^\mu{}_{\nu\rho} = \hat\Gamma^\mu{}_{\nu\rho} + \partial_\rho \alpha \, \delta^\mu_\nu\,.\label{eq:SpecificProj}
\ee
This kind of transformation was first noticed as an invariance of the metric-affine variational approach to the Einstein-Hilbert action in \cite{Einstein:1955ez}. The potential  link with the Weyl theory and with generic gauge theories was soon spotted in \cite{Bergmann:1955}  under the name of $\lambda$-transformation. In modern literature, \eqref{eq:SpecificProj} is usually indicated as an integrable projective transformation (IPT), being a particular case of the more general projective transformation (PT) \cite{Eisenhart:1927} that has been proven to be instrumental in removing ghosts from a large class of modified gravity theories \cite{BeltranJimenez:2019acz}. In practice, a PT makes manifest the link between torsion and nonmetricity, realizing the so-called torsion/nonmetricity duality \cite{Iosifidis:2018zjj}. However, it is not always possible to trade torsion for nonmetricity as in a Weyl geometry since this mechanism eventually depends on the dynamics implied by the specific theory in a generic metric-affine framework. In this respect, pure GR has a special place since absence of nonmetricity implies a contemporary absence of torsion \cite{DadPons}. Indeed, solving the connection field equations, one finds that both non-metricity and torsion depend on an arbitrary vector that can be set to zero applying a PT, thus recovering Riemannian geometry. Otherwise stated, this is a reminder that the Christoffel symbol is a solution of the field equations derived from the variation of the connection so that, applying a PT to the Riemannian solution, necessarily turns on torsion and nonmetricity at the same time.

It is worth reminding that the affine notion of parallelism is preserved under IPT and, most remarkably, the Riemann tensor itself is left invariant. Consequently, IPT can be used to unveil a natural way to gauge the conformal symmetry in metric-affine gravities, using the subtle interplay between them, diffeomorphisms and local scale transformations. This promising framework is especially attractive if one considers that conformal symmetry plays a prominent role in many proposals of  high-energy Physics and Cosmology beyond the Standard Model, related to fundamental problems like Dark Matter/Dark Energy, the Cosmological Constant, Inflation, Black-Hole physics and, of course, String Theory and  Gauge/Gravity correspondence. Furthermore, exploring dynamical geometries that generalize the Riemannian one is largely motivated by the search for a consistent and predictive quantum theory of gravity. A well known quantum theory of gravity at our disposal is (Super)string Theory \cite{Green:2012oqa,Ortin:2015hya}, that largely predicts or, better, implies, even in its effective low energy manifestations, several extensions of General Relativity, at least at energy scales where this classical theory appears to be inadequate. Important progress in the understanding of quantum geometries has also been achieved within the framework of Loop Quantum Gravity \cite{LQG1,LQG2}, whose symmetry-reduced cosmological realizations \cite{LQC} can also be interpreted as extensions of classical GR  \cite{Olmo:2008nf}. Therefore, finding a way to include conformal invariance in any modified gravity is a promising way to make a step further to find a quantum description of the gravitational interaction.
 
In this paper, motivated by the many attractive features of  gravity theories that exhibit conformal invariance \cite{FradkinVilko,Floreanini:1994yp, Codello:2012sn,Shapiro:2001rz, Iosifidis:2018zwo,Alvarez:2019hxb}, we study a large class of metric-affine gravity theories where the IPT plays the role of the conformal gauge transformation with the trace of the torsion identified as the gauge field for the conformal transformations. Assuming that the gauge field associated to the conformal symmetry is dynamical, the family of gravity theories proposed in this work naturally includes a spontaneous symmetry breaking of the conformal symmetry, following ideas largely discussed in literature (see, for instance, \cite{Smolin:1979uz}). 
We show that it is possible to find a general solution of the field equations associated to the connection if we limit ourselves to the particular case of theories that depend on the symmetric part of the Ricci tensor, the so called Ricci-Based gravity theories (RBG) \cite{Olmo:2022rhf}). The resulting on-shell theory provides a generalization of previous attempts to include conformal invariance in the Einstein-Cartan approach to gravity \cite{Obukhov:1982zn,Moon:2009zq}, recovering the role of the torsion trace as the gauge field associated to conformal symmetry.

The paper is organized as follows: in Section I we review how PT in combination with IPT of the connection correspond to reparametrizations of the related autoparallel curves, giving rise to an equivalence relation among metric-affine connections. 
In Section II we define the hidden gauge conformal symmetry, identifying  the torsion trace as the related gauge field and analyzing the corresponding transformations of the other fields. The important property resides in the fact that the gauge field is internal to the affine connection, avoiding the introduction of additional degrees of freedom.  In Section III we explore a large class of theories gauged with respect to the conformal symmetry in the metric-affine approach. In particular, we describe their Lagrangeans together with some general properties. Section IV is the most important part of the paper. It contains the solution of the equation for the connection. For Ricci-based gravities, it has a universal form leading to a specific metric-affine geometry. A key role is played by the torsion trace vector, that can also be identified with the on-shell gauge field of the Weyl conformal symmetry. The crucial tool at work allowing to find solutions homologous to those of the pure gravity case is a certain factorization of the equations of motion. In Sections V and VI we present some simple and representative examples of our proposal. Finally, Section VII contains our conclusions and a brief discussion about possible future developments.


\section{Projective transformations and consistency of parallel transport \label{eq:ProjTransfGeod}}


Projective transformations have the property to define  classes of connections that leave invariant the parallel transport in a metric-affine manifold \cite{Weyl:1919fi}. In order to review how this property can be proven, consider a smooth curve $t \rightarrow \gamma(t)=\{x^\mu(t)\}$,  parametrized in terms of $t$ on a (connected) manifold. An arbitrary vector $v^\mu$ is said to be parallel-transported along the curve $\gamma(t)$ if its components do not covariantly change along the curve\footnote{
 The covariant differential of a vector field in the natural frame is, in terms of differential forms (see, {\it e.g.} \cite{Castellani:1991et}),
\be
\nabla v^\mu = d v^\mu + \Gamma^\mu{}_\nu v^\nu\, .
\ee
It can be expanded along an holonomic basis as 
\be
\nabla_\nu v^\mu dx^\nu = \left(\partial_\lambda v^\mu + \Gamma^\mu{}_{\nu\lambda} v^\nu\right)dx^\lambda\, .
\ee
Thus, adopting these conventions, we prefer the following definition, in components, of the covariant derivative 
\be
\nabla_\lambda v^\mu = \partial_\lambda v^\mu + \Gamma^\mu{}_{\nu\lambda} v^\nu\,,
\ee
where the lower indices of the connection are switched with respect to most of the standard literature.}
\be
\frac{d \gamma^\mu}{d t}\nabla_\mu v^\nu = \frac{d v^\nu}{d t} + \Gamma^\nu{}_{\rho\mu} v^\rho \frac{d \gamma^\mu}{d t} = 0\,,\label{eq:PartParTransp}
\ee
where $\frac{d \gamma^\mu}{d t}$ is the tangent vector along the curve. In absence of a metric to define a notion of distance, the condition \eqref{eq:PartParTransp} can be slightly weakened assuming that the parallel transported vector should be proportional to itself, namely
\be
 \frac{d v^\nu}{d t} + \Gamma^\nu{}_{\rho\mu} v^\rho \frac{d \gamma^\mu}{d t}  = f(t) v^\nu\,,\label{eq:PartParTranspGener}
\ee
where $f(t)$ is an arbitrary function along the curve. It has been known for a long time that a projective transformation is an invariance of the notion of parallelism in a metric-affine theory.  The argument that justifies this statement was first proposed in \cite{Eisenhart:1927} and works as follows.
Multiplying by $v^\sigma$ equation \eqref{eq:PartParTranspGener} and antisymmetrizing, one can get rid of the extra term on the rhs, thus finding the characterization of parallel vectors along any curve
\be
v^{[\sigma}\frac{d v^{\nu]}}{d t} + \Gamma^{[\nu}{}_{\rho\mu} v^{\sigma]}v^\rho \frac{d \gamma^\mu}{d t}=0\,.\label{eq:ParTransp}
\ee
It is natural to try to address the question of whether it is possible that different connections share the same parallelism. Subtracting Eq.(\ref{eq:ParTransp}) for two connections ${\Gamma^\prime}$ and $\Gamma$, it amounts to require that they must satisfy the equation
\be
\delta_{\nu\sigma}^{\alpha\beta}\left(\Gamma^\nu{}_{\rho\mu}-\Gamma^{\prime\nu}{}_{\rho\mu}\right)v^\sigma v^\rho \frac{d\gamma^\mu}{dt} = 0\, .
\ee 
The solution to this equation implies that the two connections must be related by 
\be
\Gamma^{\prime\nu}{}_{\rho\mu} = \Gamma^\nu{}_{\rho\mu} + \delta^\nu_\rho A_\mu\, \,  ,\label{eq:ProjTransf}
\ee
where $A_\mu$ is an arbitrary vector. This relation defines a projective transformation (PT) of the connection. \\

Turning back to Eq.(\ref{eq:PartParTranspGener}), since parallel transport should be innocuous to the choice of $f(t)$, we expect that some mechanism should be able to cancel the right-hand side of \eqref{eq:PartParTranspGener}. 
Given that for an arbitrary $v^\mu$ a reparametrization $t\rightarrow  \tau(t)$ does not help, because it simply rescales the function $f(t)\rightarrow f(t)dt/d \tau$, the only option left is a PT, which changes the generalized parallel transport equation (\ref{eq:PartParTranspGener}) as follows
\be
 \frac{d v^\nu}{d t} + \Gamma^{\prime\nu}{}_{\rho\mu} v^\rho \frac{d \gamma^\mu}{d t} = \left(f(t) - A_\mu\frac{d \gamma^\mu}{d t}\right) v^\nu\,,\label{eq:PartParTranspGenerX}
\ee
so that the usual parallel transport equation (\ref{eq:PartParTransp}) is recovered when the 1-forms $f(t)dt$ and $A_\mu d\gamma^\mu$ are identified: 
\be
A_\mu d\gamma^\mu= {f(t)\,dt}\ 
\ee
On the other hand, in a metric-affine geometry, a rescaling of the vector $v^\mu\rightarrow e^{-\alpha(t)} v^\mu$  should not affect the parallel-transport condition. Demanding that the rescaled vector is still a solution of the parallel transport equation (\ref{eq:PartParTranspGener}), we find that it satisfies
\be
\frac{d v^\nu}{d t} + \Gamma^\nu{}_{\rho\mu} v^\rho \frac{d \gamma^\mu}{d t} = \left(f(t)+ \dfrac{d\alpha}{dt}\right)v^\nu \, .\label{eq:GenParTransp}
\ee
The right-hand side of this equation can be set to zero by a PT of the form 
\be
A_\mu d\gamma^\mu= {f(t)\,dt} + d\alpha \ ,\label{eq:SpecPT} 
\ee
{where the last term is an exact differential. This means that $A_\mu$ can be split as the sum of a PT plus an IPT, $A_\mu=B_\mu+\partial_\mu \alpha$, where the IPT takes care of cancelling the effect of the scale transformation. Thus, if we want to impose the scaling invariance of the parallel transport equation, we must associate with each scale transformation an IPT of the connection. }

Let us now focus on the autoparallel equation, which is the case when $v^\mu$ is replaced by $v^\mu=d\gamma^\mu/dt$ in Eq.(\ref{eq:PartParTranspGener}). This equation defines the notion of {\it autoparallel path}, namely,  a curve whose tangent vector is parallel-transported along its own direction according to the following equation
\be
\frac{d^2 \gamma^\nu}{d t^2} + \Gamma^\nu{}_{\rho\mu} \frac{d\gamma^\rho}{dt} \frac{d \gamma^\mu}{d t} = f(t)\frac{d\gamma^\nu}{dt} \ .
\ee

As mentioned above, reparametrization is not an invariance of the parallel transport equation because it rescales the function $f(t)\rightarrow f(t)dt/d \tau$. In the autoparallel equation, a reparametrization $t\rightarrow \tau(t)$ adds an additional piece on the right-hand side, leading to 
\be
\frac{d^2 \gamma^\nu}{d \tau^2} + \Gamma^\nu{}_{\rho\mu} \frac{d\gamma^\rho}{d\tau} \frac{d \gamma^\mu}{d \tau} = \left(f(t)-\frac{d^2\tau}{dt^2}\right)\left(\frac{dt}{d\tau}\right)\frac{d\gamma^\nu}{d\tau}\, .\label{eq:RepTranspEq}
\ee
Considering a scaling transformation $\frac{d\gamma^\rho}{d\tau} \rightarrow e^{-\alpha(\tau)}\frac{d\gamma^\rho}{d\tau} $ and a PT, the resulting equation takes the form
\be
\frac{d^2 \gamma^\nu}{d \tau^2} + \Gamma^\nu{}_{\rho\mu} \frac{d\gamma^\rho}{d\tau} \frac{d \gamma^\mu}{d \tau} = \left[\left(f(\tau)-\frac{d^2\tau}{dt^2}\right)\left(\frac{dt}{d\tau}\right)+\frac{d\alpha}{d\tau}-A_\mu\frac{d \gamma^\mu}{d \tau}\right]\frac{d\gamma^\nu}{d\tau} \ ,
\ee
and the condition to set the right-hand side to zero becomes
\be
A_\mu d\gamma^\mu= {f(t)\,dt} + d\alpha -d\left(\frac{d\tau}{dt}\right)\ ,\label{eq:SpecPTgeo} 
\ee
leading to the standard autoparallel equation.
This result tells us that since a reparametrization requires a twice differentiable function $\tau(t)$, a scaling $\alpha(\tau)$ can compensate a reparametrization if $\alpha(\tau)$ is differentiable at least once, so that the relation $\alpha=d\tau/dt$ can indeed be satisfied. Any combination of scaling and reparametrization, $\beta(\tau)\equiv\alpha(\tau)-  d\tau/dt$ can be compensated by an IPT, while the continuous but non-differentiable part of $f(t)$ always requires a full PT to be gauged away. 

The interplay between scaling, reparametrization, and IPT makes it clear that IPT symmetry is necessary to restore parametrization invariance and independence of the scaling of lengths for the standard parallel transport and autoparallel equations.\\


\section{Gauge symmetry from Integrable Projective Transformations  \label{eq:GaugeSymm}}


As shown in Section I, IPT is mandatory in order to have a consistent parallel transport on a metric-affine spacetime. It is therefore crucial to discuss its meaning relatively to the scale transformations applied to other fields. To this end, let us notice that PTs play a prominent role when a gravity theory is formulated using the metric-affine approach. In fact, given the independence between metric and connection, it is possible to perform a transformation of the connection without affecting the metric, as far as the theory remains off-shell. The transformation of the Riemann tensor induced by the projective transformation in \eqref{eq:ProjTransf} is\footnote{In our conventions the components of the curvature Riemann two form are defined as
$
R^\mu{}_\nu = 2R^\mu{}_{\nu\rho\sigma}dx^\rho\wedge dx^\sigma = d\Gamma^\mu{}_\nu + \Gamma^\mu{}_\lambda\wedge \Gamma^\lambda{}_\nu\,.
$
Thus, the ordering of the remaining lower indices is 
\be
R^\mu{}_{\nu\rho\sigma}(\Gamma) = 2\left(\partial_{[\rho|}\Gamma^\mu{}_{\nu|\sigma]}+\Gamma^\mu{}_{\lambda[\rho|}\Gamma^\lambda{}_{\nu|\sigma]}\right)\,.
\ee
}
\be
R^\rho{}_{\sigma\mu\nu}({\Gamma^\prime})=R^\rho{}_{\sigma\mu\nu}(\Gamma)+2\partial_{[\mu}A_{\nu]}\delta^\rho_{\sigma}\,. \label{RiemannTransf}
\ee
It shows that, in the particular case of an IPT \eqref{eq:SpecificProj}, namely when $\ A_{\mu}= \partial_{\mu} \alpha(x)$, 
the Riemann tensor is itself invariant. It follows that the Riemann tensor can be directly used to build Lagrangian that are invariant under an IPT, paving the way to unveil a (classical) hidden gauge symmetry that pertains to any gravity theory based on the curvature tensor. Indeed, this symmetry is realized by the following scaling transformation of (primary) $(m,n)$ tensor field \footnote{
Since the coordinates are not assumed to scale, there is no change in the argument of the tensor.} 
\be
{V}^{\prime\mu_1...\mu_m}{}_{\nu_1...\nu_n} = e^{(n-m)\alpha(x)}V^{\mu_1...\mu_m}{}_{\nu_1...\nu_n}\,.\label{eq:TensFieldTransf}
\ee
The gauge field of this transformation is already embedded within the affine connection of the ordinary covariant derivative in such a way that the derivative of a primary field transforms as the primary field itself, namely 
\be
{\nabla^\prime}_\rho  {V}^{\prime\mu_1...\mu_m}{}_{\nu_1...\nu_n} = e^{(n-m)\alpha(x)} {\nabla}_\rho V^{\mu_1...\mu_m}{}_{\nu_1...\nu_n}\,,\label{eq:ScaleTransf}
\ee
provided that the affine connection gauge-transforms according to the IPT in eq.\eqref{eq:SpecificProj}, exactly compensating the scaling weight $(n-m)$, with $n$ and $m$ the number of contravariant and covariant indices of the primary field, respectively. To single out the part of the connection that is responsible for the scaling gauge transformation, it should be noticed that it is always possible to decompose the connection as $\Gamma^\mu{}_{\nu\rho} = \tilde\Gamma^\mu{}_{\nu\rho} + 2/3 \ \delta^\mu_\nu S_\rho$ where, as seen, $S_\rho\equiv S^\lambda{}_{\lambda\rho}$ is the torsion trace. $\tilde\Gamma^\mu{}_{\nu\rho}$ is the so called Schr\"odinger ``star affinity'' \cite{Schrodinger:2011gqa} and it has two virtues: first it is invariant under PT, second, its torsion exhibits vanishing trace.  As a consequence, by defining the gauge field associated to the \sout{conformal gauge} local scaling symmetry to be $B_{\mu} \equiv 2 S_{\mu}/3$, its gauge transformations follows from \eqref{eq:SpecificProj} and reads\footnote{Using directly the trace of the whole connection $\Gamma_\mu \equiv \Gamma^{\lambda}{}_{\lambda\mu}$ as the gauge field is not viable, since $\Gamma_\mu$ is not a vector under diffeomorphisms, due to the inhomogeneous part of its transformation.}
\be
B^\prime_{\mu} = B_{\mu} + \partial_\mu\alpha(x) \,.\label{eq:GaugeTransfGamma}
\ee
Moreover, it should be noticed that the gauge field $B_{\mu}$ is one of the vectorial components of the affine connection \cite{Shapiro:2001rz,Capozziello:2001mq}.
However, the definition of the covariant derivative of the ``descendants'' fields, obtained by derivatives of the corresponding primary fields, demands the explicit introduction of the gauge field $B_\mu$. In formulas,
\be
{\cal{D}}_\rho{\cal{D}}^{(p)}_{\lambda_1\cdots\lambda_p}V^{\mu_1...\mu_m}{}_{\nu_1...\nu_n}\equiv \left(\nabla_\rho + p\, B_\rho\right){\cal{D}}^{(p)}_{\lambda_1\cdots\lambda_p}V^{\mu_1...\mu_m}{}_{\nu_1...\nu_n}\,,
\ee
where ${\cal{D}}^{(p)}_{\lambda_1\cdots\lambda_p} = {\cal{D}}_{\lambda_1}\cdots {\cal{D}}_{\lambda_p}$.
The fields that transform according to the mentioned rules are the building blocks of scaling invariant objects that enters a conformal gravity theory based on the Riemann curvature. As already mentioned, this kind of gauge symmetry was first noticed by Bergmann generalizing Einstein's $\lambda$ (constant) transformations, but surprisingly not totally appreciated in literature (see, however, \cite{Julia1,Julia2,DadPons,Iosifidis:2018zwo}). 

It is very important to observe that the corresponding gauge field $B_{\mu}$ associated to the conformal symmetry is embedded in the affine connection. Indeed, as seen in the introduction, it is easy to demonstrate by decomposition in irreducible components that a generic affine connection contains two vector fields, the torsion trace and the Weyl vector \cite{Shapiro:2001rz}. In particular, we would like to use the former as the gauge field, because it comes out to be the natural choice in order to solve the dynamics of the connection in a  Palatini approach\footnote{Of course, the Weyl vector can be used as well. See {\it e.g.} \cite{Ghilencea:2019jux}}.

It is worth to stress that the existence of this (hidden) gauge symmetry is consistent with the fact that four degrees of freedom of the connection are not fixed by the connection field equations in a metric-affine formalism, as emphasized in \cite{Afonso:2017bxr}, where this freedom has been used to set the (dynamical part of the) torsion to zero. 


\section{Conformal metric-affine modified gravities}


The local scaling symmetry described in Section \eqref{eq:GaugeSymm} is instrumental to the introduction  of (classical) conformal invariant gravity theories in the framework of metric-affine formalism. In fact, the transformation \eqref{eq:TensFieldTransf} applied to the metric, is the usual conformal (or Weyl) transformation of Eq.\eqref{eq:WeylTransf}\footnote{
Since the coordinates are not assumed to change, note that we are considering an "active" transformation. Alternatively, one can interpret the Weyl transformation as a scale transformation acting on the space-time coordinates \cite{Iorio:1996ad} 
\be
\tilde{x}^{\mu}=e^{\alpha(x)}x^{\mu}\, , \qquad \tilde{x}^{\mu} g_{\mu\nu}\tilde{x}^\nu = e^{2\alpha(x)}x^{\mu} g_{\mu\nu}x^\nu \, ,
\ee 
with the (naive) conformal weights of the matter fields fixed by \eqref{eq:TensFieldTransf}. 
}.
As already said, the price to pay to gain a conformal symmetry is to assume that the (naive) scaling weights of the matter fields follow from their tensorial representation properties under diffeomorphisms, a quite unusual property.\footnote{For fermions, the scaling weights can be deduced by the correct scaling of the kinetic terms in the Lagrangian, as for the engineering dimension. }
Remarkably, as observed before, the covariant derivative of the metric transforms properly (by construction) under Weyl rescalings,
\be
{\nabla^\prime}_\mu {g^\prime}_{\nu\rho} = e^{2\alpha(x)} \, {\nabla}_\mu  g_{\nu\rho}\,,
\ee
provided that the connection transforms following \eqref{eq:SpecificProj}. In particular, the metricity condition is obviously preserved by conformal rescalings. This setting differs from the original Weyl proposal in \cite{Weyl:1918ib} (and the vast literature based on it), because in our proposal the gauge field   associated to the conformal transformations is already included in the geometry, being identified with the trace of the torsion. Moreover, this is also different from the Einstein-Cartan formalism, where the metricity condition is an apriori assumption while in our case the relation between the connection and the metric is dictated dynamically by a variational principle. The intuition that a fundamental theory of gravity should be local scale invariant at short distances and that a spontaneous symmetry breaking of this invariance could help in finding the right setting for a quantum theory of gravity is a longstanding idea \cite{Smolin:1979uz,Floreanini:1994yp,FradkinVilko} that fostered a number of possible theories based on Weyl Geometry \cite{Wheeler:2018rjb} or on Einstein-Cartan formalism \cite{Hehl:1976kj, Blagojevic:2012bc}. Throwing the IPT into the game, it becomes apparent that a more natural framework to set up a conformal theory of gravity is the metric-affine geometry, thus generalizing previous approaches. 
In the metric-affine formalism, the non-minimal couplings  can be adapted to extend conformal invariance to a whole class of theories using, as customary, a compensating scalar field (a dilaton) $\Phi(x)$ \cite{Bekenstein:1980jw}. Let us start by discussing the gravity sector, that we assume of the form
\be
S_G = \int{d^4x\,\sqrt{-g} \,{\cal L}_G\left[\Phi,g_{\mu\nu},R^\lambda{}_{\mu\nu\rho}(\Gamma)\right]}\,,\label{eq:GravLag}
\ee
where the Lagrangian scalar density ${\cal L}_G\left[\Phi,g_{\mu\nu}R^\lambda{}_{\mu\nu\rho}(\Gamma)\right]$ is an arbitrary function of a collection of terms depending on diffeomorphic invariant contractions of the Riemann tensor. Although the Riemann tensor is invariant under IPT, invariants built with its contractions are not. Indeed,  the presence of a certain number of metric terms forces the introduction of the scalar field $\Phi$ in order to compensate the conformal rescalings.  This is a property of all the terms in the Lagrangean. In order to get a conformal invariant kinetic term for the compensating scalar $\Phi$, it is common to assume that it scales as 
\be
{\Phi^\prime}=e^{- \alpha(x)}\Phi\,,\label{eq:PhiConfTrans}
\ee
under (gauge) conformal transformations. Since the compensator field, as defined, does not behave as a primary field being charged with respect to the Weyl scaling, the covariant derivative results\footnote{It is clear that the compensator/BD field does not follow the general rule \eqref{eq:TensFieldTransf} so that the definition of the covariant derivatives has to be modified as follows
${\cal{D}}_\rho{\cal{D}}^{(p)}_{\lambda_1\cdots\lambda_p}\Phi\equiv \left[\nabla_\rho + (p+1)\, B_\rho\right]{\cal{D}}^{(p)}_{\lambda_1\cdots\lambda_p}\Phi$
.}
\be
{\cal{D}_\mu} \Phi = \left(\partial_\mu + B_\mu\right) \Phi = \left(\partial_\mu + \frac{2}{3} \, S_\mu\right) \Phi\,,\label{eq:CovDerPhi}
\ee
where the weight under scaling transformation is consistent with \eqref{eq:GaugeTransfGamma}. 
Taking into account the scaling of the measure in the integral, each term must contain a factor $\Phi^{4-2k}$, with $k$ counting the net number of contravariant metrics appearing in the contractions of each diffeomorphic invariant term. For instance, in the Brans-Dicke (BD) theory, where the compensator coincides with the (BD) scalar field, $\Phi$ is non-minimally coupled to the scalar curvature with an action $S_{BD} = \int{ d^4 x \sqrt{-g} \,  \Phi^2\, R}$.  According to these prescriptions, it follows that the gravity Lagrangian scales as 
\be
{\cal L}_G\left[{\Phi^\prime},{g^\prime}_{\mu\nu},R^\lambda{}_{\mu\nu\rho}({\Gamma^\prime})\right] = e^{-4\alpha(x)}{\cal L}_G\left[\Phi, g_{\mu\nu},R^\lambda{}_{\mu\nu\rho}(\Gamma)\right]\,.\label{eq:LagConfTransf}
\ee
Demanding conformal invariance restricts the choice of a kinetic term for the Brans-Dicke field to the standard form. Furthermore, the gauge field associated to the conformal symmetry can be promoted to a dynamical field as well. Introducing the field strength $F_{\mu\nu}\equiv 2\partial_{[\mu}B_{\nu]} = \frac{4}{3}\partial_{[\mu}S_{\nu]}$ of the gauge field, we can define a ``St\"uckelberg sector''
\be
S_S =  \int{d^4x\,\sqrt{-g} \, \left( \xi_1 \,  {\cal{D}_\mu}\Phi{\cal{D}^\mu}\Phi \, - \, \frac{\xi_2}{4} \, F_{\mu\nu}F^{\mu\nu}\right)}\,.\label{eq:Stuckelberg}
\ee
It should be noticed that $S_S$ contains a gauge invariant mass term \`a la St\"uckelberg, with $\Phi$ playing exactly the role of a St\"uckelberg field, as neatly evidentiated by recasting \eqref{eq:Stuckelberg} as
\be
S_{S} = \int{d^4x\,\sqrt{-g} \, \left[ \xi_1 \left(\partial_\mu \Phi +B_\mu \Phi\right)\left(\partial^\mu \Phi + B^\mu \Phi \right) - \frac{\xi_2}{4} \,  F_{\mu\nu}F^{\mu\nu} \right]}\,.
\ee
The class of theories of interest for this work is given by the sum of the gravity and St\"uckelberg sectors, supplemented possibly  by additional matter fields and leading to the following action
\be
S=S_G+S_S+S_m\,,\label{eq:Action}
\ee
where
\be
S_m = \int{d^4x\,\sqrt{-g}\, {\cal L}_m\left(g_{\mu\nu},\Phi,{\cal{D}_\mu}\Phi,\psi, \partial_{\mu}\psi \right)}\,.
\label{eq:matteraction}
\ee
The dependence of the matter sector in \eqref{eq:matteraction} is consistent with the fact that we assume couplings between generic matter fields $\psi$ and gravity not involving the whole connection but just the torsion trace proportional to the gauge vector $B_\mu$\footnote{We consider only spin zero and spin 1 matter fields. The fermion case will be dealt with in a different paper.}.
In principle, a kinetic term for the gauge field $B_\mu$ could stem from the higher derivative terms of the gravity Lagrangian. If this is the case, then the coefficient $\xi_2$ would be shifted accordingly.

\vspace{0.5cm}

Our proposal provides a simple prescription to define conformal invariant metric-affine theories of gravity. As mentioned, the meaning of the St\"uckelberg field can be exposed by choosing an appropriate gauge.  Indeed, in the (trivial) choice $\Phi=1$, one recovers the gravity sector in the Einstein frame, coupled to the matter fields and to a massive (Proca) field $B_{\mu}$:  
\be
S=\int{d^4x\,\sqrt{- g} {\cal L}_G\left(g_{\mu\nu},R^\alpha{}_{\beta\mu\nu}(\Gamma)\right)} + \int{d^4x\,\sqrt{- g}\left(- \frac{\xi_2}{4} F_{\mu\nu} F^{\mu\nu} + \xi_1 B_\mu B^\mu  + {\cal L}_m\left(g_{\mu\nu},B_{\mu},\psi,\partial_{\mu}\psi\right) \right)}\,.
\label{eq:actionproca}\ee
Finally, a mass scale is naturally introduced for the gauge field $B_{\mu}$ as a consequence of its dynamics, in association to the conformal symmetry.


\section{Solving the connection field equation }


The field equations associated to the variation of the connection for the action \eqref{eq:Action} can be found following the steps of \cite{Afonso:2017bxr} with the additional contribution of the kinetic term and interactions of the gauge field $B_{\mu}$, thus generalizing the minimally coupled case where it is assumed that $\dfrac{\delta S_m}{\delta \Gamma^\alpha{}_{\beta\nu}}=0$. After some algebra, the field equations can be written as follows:
\begin{align}
&\frac{1}{\sqrt{-g}}\nabla_\nu \left(\sqrt{-g}{P_\alpha}^{\beta\nu\mu} \right) + S^\mu{}_{\rho\nu}{P_\alpha}^{\beta\rho\nu} - 2S_\nu{P_\alpha}^{\beta\mu\nu} \nonumber\\ 
&= -\frac{1}{3 \sqrt{-g}} \left[ 2 \sqrt{-g}\,\xi_1\, \Phi {\cal{D}}^{\nu} \Phi + \xi_2\,\partial_\rho\left(\sqrt{-g}F^{\rho\nu}\right) +  \sqrt{-g}\,\frac{\partial {\cal L}_m}{\partial B_\nu}\right]\delta^{\beta\mu}_{\alpha\nu}\,,\label{eq:ConnectionFieldEquation}
\end{align}
where the tensor ${P_\gamma}^{\beta\mu\nu} = \frac{\partial {\cal L}_g}{\partial {R^\gamma}_{\beta\mu\nu}}$ has been introduced and {the torsion tensor is identified as $S^\sigma{}_{\mu\nu} \equiv \Gamma^\sigma{}_{[\mu\nu]}$ {together with its torsion trace $S_\mu\equiv S^\lambda{}_{\lambda\mu}$}. The field equation that results from taking the trace in $\gamma$ and $\beta$, can be used to decouple the gravitational sector from the non-gravitational one reducing eqs. \eqref{eq:ConnectionFieldEquation} to
\be
\frac{1}{\sqrt{-g}}\nabla_\nu \left(\sqrt{-g}~{U_\alpha}^{\beta\mu\nu} \right) - S^\nu{}_{\rho\sigma}{U_\alpha}^{\beta\rho\sigma} + 2S_{\nu}{U_\alpha}^{\beta\mu\nu} = 0\,,
\ee
\begin{align}
\xi_2 \,\partial_\mu\left(\sqrt{-g}F^{\mu\nu}\right) =  -2\sqrt{-g}\left[\frac{1}{\sqrt{-g}}\nabla_\mu \left(\sqrt{-g}P^{\mu\nu} \right) + S^\nu{}_{\rho\sigma}P^{\rho\sigma} + 2S_{\mu}P^{\mu\nu} + \xi_1\Phi {\cal{D}}^{\nu} \Phi + \dfrac12 \frac{\partial {\cal L}_m}{\partial B_\nu} \right]
\,,
\end{align}
where we defined $P^{\mu\nu}\equiv \delta^\alpha_\beta {P_\alpha}^{\beta\mu\nu}$ and the tensor ${U_\alpha}^{\beta\mu\nu}\equiv {P_\alpha}^{\beta\mu\nu}-\dfrac23 P^{\mu\rho}\delta_{\alpha\rho}^{\beta\nu}$.

If the gravity sector is a function of contractions of the Ricci tensor $R_{\mu\nu} = \delta^{\rho}_\sigma {R^\sigma}_{\mu\rho\nu}$,  we can simplify the equations above writing the P-tensor as ${P_\alpha}^{\beta\mu\nu} = Z^{\beta\rho}\delta^{\mu\nu}_{\alpha\rho}$, where now $Z^{\mu\nu}\equiv\dfrac{\partial{\cal L}_G}{\partial R_{\mu\nu}}$. In turn, the U-tensor becomes 
\be
{U_\alpha}^{\beta\mu\nu} = Z^{\beta\rho} \delta^{\mu\nu}_{\alpha\rho}-\frac23 Z^{[\mu\rho]}\delta_{\alpha\rho}^{\beta\nu}\,.
\ee
Once again, the so called RBG have a privileged role. Indeed, $ Z^{\mu\nu}= Z^{(\mu\nu)}$ and 
\be
{U_\alpha}^{\beta\mu\nu} = Z^{\beta\rho} \delta^{\mu\nu}_{\alpha\rho}
\ee
so that, despite the presence of couplings between some degrees of freedom of the connection and matter fields, the field equations for the connection reduce to those of any minimally coupled theory of gravity, namely
\be
\frac{1}{\sqrt{-g}}\nabla_\nu\left(\sqrt{-g} {Z}^{\beta\lambda}\right) \delta^{\mu\nu}_{\alpha\lambda} + S^\mu{}_{\alpha\lambda}{Z}^{\beta\lambda} + 2 S_\nu{Z}^{\beta\rho}\delta^{\mu\nu}_{\alpha\rho} = 0\, .\label{eq:ConnFieldEq}
\ee
Moreover, since for any RBG theory the Z-tensor is symmetric, it follows that $P^{\mu\nu}=0$ so that the remaining field equations in \eqref{eq:ConnectionFieldEquation} reduce to 
\be
\xi_2\,\partial_\nu\left(\sqrt{-g}F^{\nu\mu}\right) = -2 \sqrt{-g}\left(\,\xi_1 \Phi {\cal{D}}^\mu{\Phi} + \,\dfrac12\frac{\partial {\cal L}_m}{\partial B_\mu}\right)\,.\label{eq:WeylDynamics}
\ee
Since we managed to reduce the field equations involving the connection to the minimally coupled case, we can follow the procedure in \cite{Orazi:2020mhb}, that is a refined version of the resolution presented in \cite{Afonso:2017bxr}, to solve them and find a general solution. To this end, we recast \eqref{eq:ConnFieldEq} as 
\be
\nabla_\mu q^{\nu\rho} = \frac23\left(S_{\mu} q^{\nu\rho} + S_{\sigma} q^{\sigma(\nu}\delta^{\rho)}_\mu\right) - 2S^{(\nu}{}_{\mu\sigma}q^{\sigma)\rho}\,,\label{eq:qConnFieldEq}
\ee
in terms of the auxiliary metric
\be
q^{\mu\nu} = \left| \det \left[(Z^{-1})_{\lambda\sigma} \, g^{\sigma\rho} \right] \right|^{\frac12} \, Z^{\mu\nu} \,,
\label{eq:defq}
\ee
provided that $Z^{\mu\nu}$ is invertible. Next, we notice that the field equations \eqref{eq:qConnFieldEq} covariantly transform under a PT. Denoting with a tilde the fields in the new frame, we are free to choose the PT in such a way that $\tilde S_\mu = 0$, that results in
\be
\tilde\Gamma^\lambda{}_{\mu\nu} = \Gamma^\lambda{}_{\mu\nu} {\color{black} -} \frac23  \delta^\lambda_\mu S_\nu \,,\label{eq:SpecProjTransf}
\ee
and reduces the field equations to
\be
\tilde\nabla_\mu q^{\nu\rho} = - 2\tilde S^{(\nu}{}_{\mu\sigma}q^{\sigma)\rho}\,,\label{eq:qConnFieldEqFin}
\ee
leading to the solution
\be
\tilde\Gamma^\lambda{}_{\mu\nu} = \frac12 q^{\lambda\sigma}\left(\partial_\mu q_{\sigma\nu} + \partial_\nu q_{\sigma\mu} - \partial_\sigma q_{\mu\nu}\right)\,.\label{eq:MetrComp}
\ee
Eq. \eqref{eq:MetrComp} justifies the definition of the auxiliary metric as the right choice to recover a Riemannian geometry in the tilded frame since the corresponding tilded connection coincides with the Christoffel symbols of the metric $q_{\mu\nu}$ (namely the Levi Civita connection) and the tilded torsion vanishes. Finally, the solution of the field equations \eqref{eq:ConnectionFieldEquation} is
\be
\Gamma^\lambda{}_{\mu\nu} = \frac12 q^{\lambda\sigma}\left(\partial_\mu q_{\sigma\nu} + \partial_\nu q_{\sigma\mu} - \partial_\sigma q_{\mu\nu}\right)+ \delta^\lambda_\mu B_{\nu}\,,
\ee
that dynamically realizes the decomposition into a conformal invariant component of the connection, namely the Christoffel symbol of $q_{\mu\nu}$, and the gauge vector $B_\mu$.
It should be noticed that \footnote{It is worth underline that we are not considering diffeomorphic equivalent frames. Moreover, this is not a conformal gauge transformation since the metric and connection undergo indipendent redefinitions.
} 
the original torsion is completely determined by the four degrees of freedom of the conformal gauge field according to
\be
S^\lambda{}_{\mu\nu} = \delta^\lambda_{[\mu} B_{\nu]}\,.
\ee

Recalling that the Ricci tensor is invariant under an IPT, from \eqref{eq:LagConfTransf} we can figure out the scaling properties of the Z-tensor under a conformal transformation, namely
\be
 Z^{\prime\mu\nu} = e^{- 4\alpha}Z^{\mu\nu}\, .
\ee
Therefore, given its definition \eqref{eq:defq}, it is easy to show that the auxiliary metric $q_{\mu\nu}$ is invariant under conformal transformations, consistent with the fact that the star affinity $\tilde\Gamma$ is conformal invariant.

It should be observed that the complete on-shell theory demands the solution of the field equations \eqref{eq:WeylDynamics}. They establish that the dynamics of the gauge vector $B_{\mu}$ is determined by interactions with matter fields. Moreover, $B_{\mu}$ is also contributing to the field equations of the metric field through its energy-momentum tensor. It follows that the presence of a dynamical gauge field associated to the conformal symmetry crucially changes the properties of the complete solution, entering non trivially in all the field equations of the theory. In particular, the field equations associated to the dilaton field $\Phi$ read
\be
\xi_1\mathring{\cal D}_\mu{\cal D}^\mu\Phi = 3\xi_1B^\mu{\cal D}_\mu \Phi+ \frac12\left(\frac{\delta {\cal L}_G}{\delta \Phi} + \frac{\delta {\cal L}_m}{\delta \Phi}\right)\label{eq:DilFieldEqs}
\ee
where the connection entering $\mathring{\cal D}$ is the Christoffel symbol of the metric $g_{\mu\nu}$. In the unitary gauge, these equations provide further constraints on the gauge field $B_\mu$ given by
\be
\xi_1\mathring{\nabla}_\mu B^\mu = \xi_1 B_\mu B^\mu + \frac12\left.\left(\frac{\delta {\cal L}_G}{\delta \Phi} + \frac{\delta {\cal L}_m}{\delta \Phi}\right)\right|_{\Phi=1}\,,
\ee 
that should be combined with \eqref{eq:WeylDynamics} and the field equations of the metric to find the ultimate solution.

The solution of the field equations involving the connection confirms that the on-shell gauge field associated to conformal symmetry is identified with the torsion vector. As a byproduct, this implies that the on-shell theory should include a non-vanishing torsion, thus confirming that the Einstein-Cartan geometry is the proper framework for a conformal gravity theory in the metric approach. In order to make a detailed study of the on-shell underlying geometry, we provide some examples related to our proposal, aiming at a more complete analysis of the solutions and the introduction of gravitational coupling to fermions in future works \cite{inprep}.


\section{Conformal metric-affine General Relativity}


The simplest instance of a gravity theory that can be uplift to its conformal generalization is given by metric-affine GR non-minimally coupled to the dilaton (CGR). It is characterized by the choice of the metric-affine generalization of the Brans-Dicke theory in the gravitational sector. The action of this theory can be obtained by specializing \eqref{eq:Action} to the form 
\be
S_{CGR} = \int{d^4x\,\sqrt{-g}\left[\frac{1}{2\kappa^2}\Phi^2\,R\left(g_{\mu\nu},\Gamma\right) + \xi_1 {\cal{D}_\mu}{\Phi\cal{D}^\mu}\Phi - \frac{\xi_2}{4}  F_{\mu\nu}F^{\mu\nu} + {\cal L}_m\left(g_{\mu\nu},\Phi,{\cal{D}_\mu}\Phi,\psi, \partial_{\mu}\psi\right) \right]}\,,\label{eq:BransDickeAff}
\ee
and describes the metric-affine conformal invariant formulation of GR non-minimally coupled to the dilaton. In the unitary gauge, it  reduces to a metric-affine GR in the Einstein frame, coupled to a massive Proca and matter fields 
\be
S_{BD} = \int{d^4x\,\sqrt{-g}\left[\frac{1}{2\kappa^2} R\left(g_{\mu\nu},\Gamma\right) - \frac{\xi_2}{4} F_{\mu\nu} F^{\mu\nu} + \xi_1 B_\mu B^\mu  + {\cal L}_m\left(g_{\mu\nu},B_{\mu},\psi, \partial_{\mu}\psi\right) \right]}\,.\label{eq:BransDickeAffgaugefix}
\ee
From the introduction of a non-minimally coupled dilaton in the gravity sector, ${\cal L}_G\left[R^\lambda{}_{\mu\nu\rho}(\Gamma)\right] = \Phi^2 R$, it follows that eq. \eqref{eq:defq} implies
\be
q_{\mu\nu} = \Phi^2\,g_{\mu\nu}\,,\label{eq:MetricMapCGR}
\ee
so that the general solution for the connection  becomes 
\be
\Gamma^\mu{}_{\nu\rho} = \frac12 g^{\mu\sigma}\left(\partial_\nu g_{\sigma\rho} + \partial_\rho g_{\sigma\nu} - \partial_\sigma g_{\nu\rho}\right) + \delta^\mu_\rho \,\partial_\nu\ln{\Phi} + \delta^\mu_\nu \,\partial_\rho\ln{\Phi} - g_{\nu\rho}g^{\mu\sigma}\partial_\sigma\ln{\Phi} + B_{\rho} \delta^\mu_{\nu}\, .\label{eq:EHSolution}
\ee
As expected, this connection does not fulfill the metricity condition since
\be
\nabla_\mu g_{\nu\rho} = -2\left(B_\mu +  \,\partial_\mu\ln{\Phi}\right)\,g_{\nu\rho}\,.\label{eq:BDNonMetricity}
\ee
Nevertheless, the relation between metrics \eqref{eq:MetricMapCGR} can be interpreted as a conformal gauge transformation with $\alpha=\ln{\Phi}$, that brings to the unitary gauge $\Phi = 1$. The connection and the metricity conditions
\be
\Gamma^\mu{}_{\nu\rho} = \frac12 g^{\mu\sigma}\left(\partial_\nu g_{\sigma\rho} + \partial_\rho g_{\sigma\nu} - \partial_\sigma g_{\nu\rho}\right) + B_{\rho} \delta^\mu_{\nu}\, ,\label{eq:EHSolutionPhi=1}
\ee
\be
\nabla_\mu g_{\nu\rho} = -2\,B_\mu \,g_{\nu\rho}\, ,  \label{eq:BDNonMetricityPhi=1}
\ee
recover the well known Einstein frame of GR for a metric-affine connection where, in absence of a dynamics for the gauge field $B_\mu$, one can always choose to set $B_\mu =0$ falling into the ordinary Riemannian geometry. However, when a kinetic term for the gauge field $B_\mu$ is included, the dynamics of the torsion vector is non-trivial,  depending eventually on the specific interactions with the matter fields according to the field equations \eqref{eq:WeylDynamics}. We conclude that turning on the dynamics of the gauge field associated to the conformal symmetry could lead to new physics driven by the choice of the matter sector. A similar phenomenon already happens in the case of nondynamical torsion trace in the Einstein-Cartan approach discussed in \cite{Karananas:2021zkl} or in supergravity \cite{Freedman:2012,Castellani:1991et}.

\vspace{0.5cm}

The ``on-shell'' conformal invariant action can be easily unveiled by replacing the expression of the connection \eqref{eq:EHSolution} into the action \eqref{eq:BransDickeAff}, that becomes 
\begin{align}\label{eq:BransDickeAffOn-Shell-g}
S_{CGR} &= \displaystyle{\int{d^4x\,\sqrt{-g}\left[\frac{1}{2\kappa^2}\left(\Phi^2\,\mathring{R} + 6\, \partial_\mu \Phi\,\partial^\mu \Phi\right) + \xi_1 \, g^{\mu\nu} \, {\cal{D}_\mu} \Phi \, {\cal{D}_\nu} \Phi - \frac{\xi_2}{4} F_{\mu\nu} \, F^{\mu\nu} + {\cal L}_m\left(g_{\mu\nu},\Phi,{\cal{D}}_\mu\Phi,\psi, \partial_{\mu}\psi\right)\right]}} \, .
\end{align}
$\mathring{R}$ is the scalar curvature of the Levi-Civita connection $\mathring{\Gamma}^\mu{}_{\nu\rho}$ depending on the metric $g_{\mu\nu}$, while $F_{\mu\nu}=2\partial_{[\mu}B_{\nu]}$. The ordinary conformal point of the non-minimally coupled scalar-tensor Einstein gravity is recovered by setting $\xi_1 = \xi_2 = 0$ in absence of a matter sector. It concides with the on-shell version of the gravity sector, 
\begin{align}
S_{CP} &= \frac{1}{2\kappa^2}\displaystyle{\int{d^4x\,\sqrt{-g}\left(\Phi^2\,\mathring{R} + 6\, \partial_\mu \Phi\,\partial^\mu \Phi\right)}}\,,
\end{align}
invariant under the conformal gauge transformations \eqref{eq:WeylTransf} and \eqref{eq:PhiConfTrans}, thanks to the exact ratio of the non-minimal coupling of the dilaton $\Phi$ to gravity.

Similar actions have been proposed by other authors as a promising conformal invariant theory in the Einstein-Cartan formalism \cite{Obukhov:1982zn,Moon:2009zq,Lucat:2016eze,Blagojevic:2012bc}. The main difference is that in our case this action has a dynamical origin that is made manifest by solving the field equations of the connection in a metric-affine approach, where the underlying geometry is not a priori established. 

\vspace{0.5cm}

An interesting way to write the on-shell action of the conformal General Relativity is achieved by keeping the auxiliary metric. It steadily follows from \eqref{eq:BransDickeAff} that, in the ``q-frame'', where the auxiliary metric dictates the geometry of space-time, the action reads
\be\label{eq:BransDickeAffOn-Shell-q}
S = \displaystyle{\int{d^4x\,\sqrt{-q}\left[\dfrac{1}{2 k^2}\,\mathring{R}_q + \xi_1 \, q^{\mu\nu} \, \left({\cal{D}_\mu} \ln{\Phi}\right)\, \left({\cal{D}_\nu} \ln{\Phi}\right) - \frac14 \, \xi_2\,q^{\mu\rho}\, q^{\nu\sigma} \, F_{\mu\nu}F_{\rho\sigma} + {\cal L}_m\left(\Phi^{-2} q_{\mu\nu},\Phi,{\cal{D}}_\mu\Phi,\psi, \partial_{\mu}\psi\right)\right]}}
\ee
where $\mathring{R}_q$ is the curvature associated to a connection that coincides with the Christoffel symbol of $q_{\mu\nu}$. It is worth reminding that all the indices are rised using the auxiliary metric. In this q-frame, the dilaton decouple from the gravity sector because of the following relation
\be
\Phi^2\sqrt{-g}g^{\mu\nu} = \sqrt{-q}q^{\mu\nu}\,.
\ee
and its $\ln\Phi$ becomes an ordinary St\"uckelberg field. Meanwhile, the only reminiscence of the conformal symmetry is enclosed in the $\mathbb{R}^+$ gauge symmetry of the massive gauge field $B_\mu$ that lost its meaning as torsion vector.

The other way around suggests that GR minimally coupIed to a non-compact St\"uckelberg sector hides a conformal symmetry if the St\"uckelberg scalar is interpreted as a logarithm of the dilaton and the gauge field becomes the torsion vector of a metric-affine Brans-Dicke gravity. However, we cannot forget that the gauge field $B_\mu$ is still constrained by its field equations \eqref{eq:WeylDynamics} that, in the q-frame version reads:
\be
\xi_2\mathring\nabla^{(q)}_\mu F^{\mu\nu} = - 2 \left[\,\frac{\xi_1}{\Phi} q^{\nu\rho}{\cal D}_\rho{\Phi} + \dfrac{1}{2\Phi^4}\,\frac{\partial {\cal L}_m}{\partial B_\nu}\right] \,,\label{eq:WeylDynamicsOn-Shell}
\ee
where the covariant derivative $\mathring\nabla^{(q)}$ carries the Christoffel symbol of the $q$-metric as connection. These field equations are coupled to the ones coming from the variation of the dilaton, namely
\be
\xi_1 \mathring{\cal D}^{(q)}_\mu{\cal D}^\mu\Phi = 4\, \xi_1 {\cal D}_\mu\Phi {\cal D}^\mu\Phi - \xi_1 B_\mu{\cal D}^\mu \Phi+ \frac{\Phi^2}{2\kappa^2}\left(\Phi\mathring{R} + \kappa^2 \frac{\partial {\cal L}_m}{\partial \Phi}\right)\,,
\ee
according to \eqref{eq:DilFieldEqs} in terms of the auxiliary metric.

Remarkably, conformal invariance of the theory is not affected by the choice of the coefficient $\xi_1$ and $\xi_2$ that can be left arbitrary. This freedom can be used to consider simplified versions of the general theory whose field equations are more manageable. For instance, the case of a massless gauge field, that corresponds to $\xi_1=0$, yields a simplified dynamics for the gauge field, encoded in the following field equations
\be
\xi_2\mathring\nabla_\mu F^{\mu\nu} = - \frac{1}{\Phi^4}\frac{\partial {\cal L}_m}{\partial B_\nu} \,.
\ee
Accordingly, the action becomes
\begin{align}
S_{0} &= \displaystyle{\int{d^4x\,\sqrt{-g}\left[\frac{1}{2\kappa^2}\left(\Phi^2\,\mathring{R} + 6\, \partial_\mu \Phi\,\partial^\mu \Phi\right) - \frac{\xi_2}{4} F_{\mu\nu} \, F^{\mu\nu} + {\cal L}_m\left(g_{\mu\nu},\Phi,{\cal{D}}_\mu\Phi,\psi,\partial_\mu\psi\right)\right]}}\,,
\end{align}
that admits known solutions in the unitary gauge. On the other hand, if we remove the dynamics of the gauge field by setting $\xi_2=0$, then the field equations for $B_\mu$ simplify as follows
\be
\xi_1\,\Phi {\cal{D}}_\mu\Phi = -\dfrac{1}{2}\,\frac{\partial {\cal L}_m}{\partial B_\nu} \,.
\ee
In absence of matter, they provide the solution
\be
B_\mu = -\partial_\mu\ln{\Phi} \ ,
\ee
and since in this case $B_\mu$ is not dynamical, this is nothing but a gauge fixing that recovers the conformal point of the simplest nonminimally coupled scalar-tensor gravity
\begin{align}
S_{ST} &= \displaystyle{
\frac{1}{2\kappa^2}\int{
d^4x\,\sqrt{-g}\left(
\Phi^2\,\mathring{R} + 6\,\partial_\mu \Phi\,\partial^\mu \Phi
\right)
}
}\,.
\end{align}
However, in this case we face a further condition that comes from the dilaton field equation \eqref{eq:DilFieldEqs} and forces the scalar curvature to be vanishing $\mathring{R}=0$.

Keeping $\xi_2 = 0$, if the matter contribution $\dfrac{\partial {\cal L}_m}{\partial B_\nu}\neq 0$, then in the unitary gauge one gets an algebraic equation for $B_\mu$ in terms of the matter fields, which adds new elements on the right-hand side of the metric field equations and can potentially lead to new phenomenology. 


\section{The conformal metric-affine Eddington-Inspired Born-Infeld theory}

A second interesting model (to our knowledge new) is the conformal invariant version of the Eddington-inspired Born Infeld gravity (EiBI) (for a review, see {\it e.g.} \cite{BeltranJimenez:2017doy}).
The action for the gravity sector, obtained by our prescription, can be written in the form
\be\label{eq:EiBI}
S_{CEiBI} = M_{P}^2 \, M_{BI}^2 \, \int{d^4x\, \left[\sqrt{- \det\left(\Phi^2 \, g_{\mu\nu} + \frac{R_{(\mu\nu)}(\Gamma)}{M_{BI}^2}\right)} - \lambda \, \sqrt{- \det\left(\Phi^2 \, g_{\mu\nu} \right)}\right]} \, ,
\ee
where $M_{BI}$ is the Born-Infeld mass scale, and there is an effective cosmological constant term, related to $\Lambda=(1-\lambda) M_{BI}^2 \Phi^2$, linked to possible vev of $\Phi$ in a spontaneously broken phase, or after gauge fixing of $\Phi$. This could be instrumental in realizing a dynamical cosmological constant that fits observational variations of the acceleration of the universe by a proper choice of the matter interactions. The action is projectively invariant, being dependent only on the symmetric part of the Ricci tensor.  When the BI mass is large, the CEiBI model corrects in the ultraviolet the non-minimal CGR of the first example with an addicional cosmological constant. 

Recalling that the field equations for the connection of any conformal generalization of an RBG can be reduced to the ones of the non-conformal case given by \eqref{eq:ConnFieldEq}, we recover the well known solution given in terms of the metric-compatible auxiliary metric 
\be
\hat{q}_{\mu\nu} = \Phi^2 \, g_{\mu\nu} + \frac{R_{(\mu\nu)}(\Gamma)}{M_{BI}^2} \,,
\ee
as can be easily deduced from \eqref{eq:defq}. By direct substitution of this relation in (\ref{eq:EiBI}), the EiBI action assumes the form
\be
S_{EiBI-q} = - \lambda \,M_{P}^2 \, M_{BI}^2 \, \int{d^4x\,  \, \left[\sqrt{- \det\left(q_{\mu\nu} - \frac{\mathring{R}_q{}_{(\mu\nu)}(\mathring{\Gamma}_q)}{M_{BI}^2} \right)} - \frac{1}{\lambda}\,\sqrt{- \det\left(q_{\mu\nu}\right)}\right]} \, ,\label{eq:qframeEiBI}
\ee
which is explicitly invariant under conformal transformations (given that the metric $q_{\mu\nu}$ is) and also under projective transformations (because $R_{(\mu\nu)}$ is). It is worth to emphasize that the St\"uckelberg and matter sectors decouple completely from the gravity part as in the CGR. We also note that the action (\ref{eq:qframeEiBI}) should not be used to derive the field equations but rather one should use the original representation (\ref{eq:EiBI}). In fact, it is easy to see that (\ref{eq:qframeEiBI}) leads to ghost-like instabilities \cite{Deser:1998rj}, while (\ref{eq:EiBI}) is totally healthy \cite{BeltranJimenez:2017doy}.


\section{Conclusions and Discussion}


In this work we have shown that in spaces with an affine structure but without a notion of distance, scale transformations can be exactly compensated by suitable combinations of projective transformations (PTs) and integrable projective transformations (IPTs), thus resulting in the form invariance of the parallel transport equations. Going beyond that motivating scenario, this relation between scalings and IPTs in metric-affine spaces has allowed us to work out a formulation of gravity theories in which IPTs  actually represent a gauge symmetry associated to a part of the affine connection. In particular, we have shown how this gauge symmetry can be naturally encoded in metric-affine gravity theories coupled to matter fields of spin zero and one, to make them (classically) conformally invariant at any couplings. This requires the introduction of a compensator dilaton field, which is charged with respect to the conformal symmetry, giving rise to a St\"uckelberg sector and providing a mass for the gauge field of the conformal symmetry. The gauge field can be identified with the trace of the torsion tensor, generalizing to a metric-affine framework other approaches based on the Einstein-Cartan formulation. The main difference between the phenomenology in the Einstein-Cartan framework \cite{Ghilencea:2021lpa} and in the metric-affine formulation that we propose, is the presence of a dynamical gauge field that deeply changes the outcome of the theory, giving rise to interesting new features that will be addressed in future works. This has been made manifest by providing solutions of the connection field equations for a large class of theories whose gravity Lagrangian is based on the symmetric part of the Ricci tensor.
In particular, we have provided the simplest example of metric affine General Relativity invariant under local dilatation symmetry, and we have also shown how to build the corresponding conformal invariant version of the EiBI model. Generically, there exists a conformal frame in which a natural notion of proper time can be defined in terms of the metric $q_{\mu\nu}$, avoiding in that way the second clock effect that precluded the acceptance of the original  Weyl's theory as physically viable \cite{Perlick,Lobo:2018zrz}. Furthermore, the introduction of a dinamical conformal gauge field sets a constraint on the four degrees of freedom of the connection that are not fixed by the field equations in the metric-affine approach, recovering consistency without breaking the projective invariance \cite{Poplawski:2007}. 
The phenomenology, implications, and extensions of our proposal still need to be further explored. In fact, much work has to be done in order to extend it to more general frameworks, like, for instance, those involving spinorial matter fields. The goal would be to understand if these models can represent the asymptotic phase of quantum versions of General Relativity at high energy and to extract possible predictions on fundamental aspects related to the extension of the Standard Model of particle physics to include (quantum) gravity, like cosmic and primordial inflation, dark matter/energy and also their implications for solutions involving compact objects. From a more formal point of view, it is worth to analyze if these models can be connected to effective field theory expansions of sensible quantum gravity theories, like in asymptotic safety proposals, or related to compactifications in the landscape of (super)string theory.

\section{Aknowledgements}

We thank F. Riccioni, A. Sagnotti, A. Salvio and R. Savelli for several interesting discussions. This work is supported in part by the Spanish Grants PID2020-116567GB-C21, funded by MCIN/AEI/10.13039/501100011033 (``ERDF A way of making Europe"), the project PROMETEO/2020/079 (Generalitat Valenciana), the project H2020-MSCA-RISE-2017 Grant FunFiCO- 777740  and by the DyConn Grant of the University of Rome ``Tor Vergata''. EO also aknowledges CNPq (Brazil), grant No.314392/2021-1, for partial financial support.

\end{document}